\documentstyle[preprint,aps]{revtex}
\begin{document}
\draft
\title{Numerical investigation of iso-spectral cavities
built from triangles
}
\author{Hua Wu and D. W. L. Sprung}
\address{
  Department of Physics and Astronomy, McMaster University\\
  Hamilton, Ontario L8S 4M1 Canada
}
\author{J. Martorell}
\address {Dept.
  d'Estructura i Constituents de la Materia, Facultat F$\acute{i}$sica,\\
   University of Barcelona, Barcelona 08028, Spain
}
\date{July 1994}
\maketitle
\begin{abstract}
We present computational approaches as alternatives to the recent
microwave cavity experiment by S. Sridhar and A. Kudrolli (Phys. Rev.
Lett. {\bf 72}, 2175 (1994)) on iso-spectral cavities built from
triangles.  A straightforward proof of iso-spectrality is given based
on the mode matching method. Our results show that the experiment is
accurate to $0.3\%$ for the first 25 states. The level statistics
resemble those of GOE when the integrable part of the spectrum is
removed.
\end{abstract}
\pacs{03.65.Ge, 05.45+b, 02.90+p }

\section{Introduction}
In a well-known paper, M. Kac\cite{K66} raised the question of whether
two isospectral plane domains must actually be isometric.  This is
popularly phrased as ``Can one hear the shape of a drum?". From the
physics viewpoint, since the density of states has an asymptotic
expansion whose coefficients depend on the area, perimeter, ... of the
cavity, it is clear that at least these properties must be common, but
this leaves open the question of whether the two must be identical in
all respects. The question has now received a definitive negative
answer. C. Gordon and collaborators\cite{GWW92} have given simple
examples constructed out of seven right triangles which are
isospectral but not isometric.

These particular shapes were subjected to experimental test by Sridhar
and Kudrolli\cite{SK94}, and found to give identical spectra to an
accuracy of $0.2\%$ on average for the first 25 eigen-energies. In
this paper we present accurate numerical results with the mode
matching and with a finite difference method. The agreement between
the two kinds of calculation ensures the accuracy of the
results. Compared to the experimental spectra, we find differences of
order $0.3\%$.

In addition a very simple proof of the iso-spectrality is obtained
with the mode matching method. The wavefunction transformation from
one cavity to the other is put into a simple form. We find that our
computed wavefunction is in good qualitative agreement with the
experimental wavefunction. The level statistics resemble those of GOE
when a subset of integrable levels are removed.

In section 2 we outline the mode-matching method for the two cavities,
and in section 3 we demonstrate iso-spectrality analytically. Section
4 contains the numerical results and conclusions.

\section{The mode matching method}
The problem we wish to solve is Helmholtz's equation
\begin{equation}
(\nabla^2+E)\Psi=0
\label{eq:1}
\end{equation}
in the interior of regions shown in Fig. 1 with Dirichlet conditions
on the boundaries.  Each cavity consists of 7 equal sided right
triangles. Let the length of a short side of a triangle be the unit
length $d$. Since the cavities are made from regular shapes, we can
divide them into smaller regions as shown in Fig. 1. We will first
concentrate on cavity one. The five smaller regions are three
triangles labelled as A, B, and E, and two squares labelled as C and
D. We shall write $\Psi_{AB}$ etc.  for wavefunction values along the
internal boundary lines separating the regions. Then the auxiliary
boundary conditions on the smaller regions are conveniently written by
expanding these wavefunctions in the Fourier series:
\begin{eqnarray}
\Psi_{AB}&=&\sum_n A_{1n} \sin\frac{n\pi x}{d}\quad,\\
\Psi_{BC}&=&\sum_n B_{1n} \sin\frac{n\pi (y-d)}{d}\quad,\\
\Psi_{CD}&=&\sum_n C_{1n} \sin\frac{n\pi (y-d)}{d}\quad,\\
\Psi_{DE}&=&\sum_n D_{1n} \sin\frac{n\pi (x-2d)}{d}\quad.
\end{eqnarray}
Here $A_{1n}\ldots D_{1n}$ are mode expansion coefficients, the
subscript 1 denotes cavity one, and the summation over $n$ is from $1$
to $N$, a value truncating the infinite series summation. We shall now
see that if these auxiliary boundary conditions are assumed to be
known, the wavefunctions in each region are easily determined as long
as the energy does not happen to coincide with an eigen-energy of the
small closed region. The wavefunction for the square region is easy to
work out, and thanks to the $45^\circ$ of the triangle, the
wavefunction for the triangle is just that of a square, plus
antisymmetrization along the diagonal line. Thus:

\FL
 \begin{eqnarray}
\Psi_A&=&\sum_n A_{1n}\Bigl[
\sin \frac{n\pi x}{d}\, \frac{ \sin\alpha_n(3d-y)}{\sin\alpha_n d}
\nonumber\\
&&\phantom{\sum_nA_{1n}}
-\sin \frac{n\pi(y-2d)}{d}\, \frac{\sin\alpha_n(d-x)}{\sin\alpha_n
d}\Bigr]\quad,
\end{eqnarray}

\FL
 \begin{eqnarray}
\Psi_B&=&\sum_nA_{1n}\Bigl[\sin \frac{n\pi x}{d}
\,\frac{\sin\alpha_n(y-d)}{\sin\alpha_n d}\nonumber\\
&&\phantom{\sum_nA_{1n}}
-\sin\frac{n\pi(2d-y)}{d}
\,\frac{\sin\alpha_n(d-x)}{\sin\alpha_n d}\Bigr]\nonumber\\
&+&\sum B_{1n}\Bigl[\sin \frac{n\pi(y-d)}{d}
\,\frac{\sin\alpha_n x}{\sin\alpha_n d}\nonumber\\
&&\phantom{\sum_nA_{1n}}
-\sin\frac{n\pi(d-x)}{d}
\,\frac{\sin\alpha_n (2d-y)}{\sin\alpha_n d}\Bigr]\quad,
\end{eqnarray}

\FL
 \begin{eqnarray}
\Psi_C&=&\sum_n B_{1n} \sin \frac{n\pi(y-d)}{d}
\,\frac{\sin\alpha_n(2d-x)}{\sin\alpha_n d}\nonumber\\
&+& \sum_n C_{1n}\sin \frac{n\pi(y-d)}{d}
\, \frac{\sin\alpha_n(x-d)}{\sin\alpha_nd}\quad,
\end{eqnarray}

\FL
 \begin{eqnarray}
\Psi_D&=&\sum_n C_{1n} \sin \frac{n\pi(y-d)}{d}
\,\frac{\sin\alpha_n(3d-x)}{\sin\alpha_n d}\nonumber\\
&+& \sum_n D_{1n}\sin \frac{n\pi(x-2d)}{d}
\,\frac{\sin\alpha_n(2d-y)}{\sin\alpha_nd}\quad,
\end{eqnarray}

\FL
 \begin{eqnarray}
\Psi_E&=&\sum_n D_{1n}\Bigl[ \sin \frac{n\pi(x-2d)}{d}
\,\frac{\sin\alpha_ny}{\sin\alpha_n d}\nonumber\\
&&\phantom{\sum_nA_{1n}}
  -\sin \frac{n\pi y}{d}\frac{\sin\alpha_n(x-2d)}{\sin\alpha_nd}\Bigr]
\quad,
\end{eqnarray}
where
 \begin{equation}
\alpha_n=\sqrt{E-\left(\frac{n\pi}{d}\right)^2}\quad.
\end{equation}
By construction, these wavefunctions are already continuous
across the boundary lines. The mode expansion coefficients are
now determined by requiring the normal derivatives to be
continuous as well. A straightforward calculation gives the
following condition:

\FL
 \begin{equation}
M_1               \left[\begin{array}{l}
                             A_1\\B_1\\C_1\\D_1
                        \end{array}
                   \right] =0 \quad,\label{eq:M1}
\end{equation}
with
\FL
 \begin{equation}
M_1=\left[\begin{array}{llll}
U-2W&\overline{W}-\overline{V}/2\;\;&0&0\\
\overline{W}-\overline{V}/2\;\;&U-W&-V/2\;\;&0\\
0&-V/2&U&W\\
0&0&W&U-\overline{W}
      \end{array}\right]\quad,
\end{equation}
where $U$,$V$,$W$ are $N\times N$ matrices with elements defined by:
 \begin{eqnarray}
U_{m,n}&=&\delta_{m,n}\alpha_n\cot\alpha_n d\quad,\\
V_{m,n}&=&\delta_{m,n}\alpha_n/\sin\alpha_n d\quad,
\end{eqnarray}
and
 \begin{equation}
W_{m,n}=\frac{(m\pi/d)(n\pi/d)}{E-(m\pi/d)^2-(n\pi/d)^2} \quad.
\end{equation}
Defining the diagonal matrix $P$ of order $N$, as $P_{n,n}=(-1)^n$, then
 \begin{equation}
\overline{W}=PWP,\quad \overline{V}=PV\quad.
\end{equation}
In Eq. (\ref{eq:M1}), $A_1$ etc. are column matrices of order $N$ with
$A_{1n}$ etc. as their elements.

Normally, one would look for non-trivial solutions to Eq.
(\ref{eq:M1}) by seeking energies for which $\det(M_1)=0$. A better
way is to diagonalize $M_1$ first. Then, when scanning in energy a
zero eigenvalue is found, this energy corresponds to an eigen-mode of
the cavity and the corresponding eigenvector gives rise to the
wavefunction. When $\det(M_1)\ne 0$, the wavefunctions, eqs. (2) to
(5), on the internal boundaries vanish and the only possible
non-trivial solutions are when the energy coincides with one of the
eigen-energies of the smaller regions themselves. Thus besides the
solutions dictated by $\det(M_1)=0$, other solutions exist at the
eigen-energies of the basic triangle, which are known
analytically. The wavefunctions in the several triangles must be in
proper phase to make the normal derivatives across the boundary lines
continuous, and that fixes the relative sign in each small region.

In a totally parallel way, the condition for finding eigen-energies
for the second cavity is (see appendix A):
\FL
 \begin{equation}
M_2               \left[\begin{array}{l}
                             A_2\\B_2\\C_2\\D_2
                        \end{array}
                   \right] =0 \label{eq:M2}
\end{equation}
with
\FL
 \begin{equation}
M_2=\left[\begin{array}{llll}
U-W&\overline{W}-\overline{V}/2\;\;&0&0\\
\overline{W}-\overline{V}/2\;\;&U-W&\overline{V}/2\;\;&W\\
0&\overline{V}/2&U-W&\overline{W}\\
0&W&\overline{W}&U-W
      \end{array}\right]\quad.
\end{equation}

The mode matching method is essentially an analytical one, though it
requires numerical diagonalization and root searching.  It yields fast
convergence with respect to the truncation parameter $N$. As will be
further discussed in the next section, even after truncation the
numerical spectra of the two cavities are identical, so that comparing
them does not give any additional check on the accuracy of the
numerical results.  Therefore, as a further check we performed a more
conventional calculation using a finite difference method for the
purpose of comparison and testing of results. The implementation is
trivial: one replaces the Laplace operator in eq.\ref{eq:1} with a
five point difference formula, and eigen-energies and wavefunctions
are obtained in one single diagonalization step.

\section{Isospectrality}
The theorems proved by Gordon {\it et al.}\cite{GWW92} ensure that the
two cavities have the same spectra. The experimental work of
Ref. \cite{SK94} entailed some error in comparing the two spectra. We
shall now discuss how well the numerical computations presented here
confirm the isospectrality theorem. With the same grid size, we find
that the finite difference method gives exactly the same spectra (up
to machine precision) for the two cavities. Thus, even for a finite
grid size, and although the method is an approximation to the real
cavities, isospectrality is always precise. The same property holds
for the mode matching method. To prove this, we notice that the
matrices $M_1$ and $M_2$ are connected by an orthogonal transformation
\begin{equation}
T=\frac{1}{\sqrt{2}} \left[\begin{array}{rrrr} 0 & 1&
0&\;P\\ 1 & 0&\;P& 0\\ 0 &-1& 0& P\\ -1& 0& P& 0
\end{array}\right]\quad,
\end{equation}
 \begin{equation}
M_1=T^tM_2T\quad. \label{eq:M1M2}
\end{equation}
This proves that the determinants of $M_1$ and $M_2$ have the
same set of zeros, and therefore produce the same spectra. In
addition, substituting Eq. (\ref{eq:M1M2}) into Eq. (\ref{eq:M1})
 \begin{equation}
M_2 T \left[\begin{array}{l}
        A_1\\B_1\\C_1\\D_1
      \end{array}\right]=0\quad.
\end{equation}
Thus the wavefunctions in the two cavities are connected by
 \begin{equation}
\left[\begin{array}{l}
        A_2\\B_2\\C_2\\D_2
      \end{array}\right]=
    T \left[\begin{array}{l}
        A_1\\B_1\\C_1\\D_1
      \end{array}\right]\quad.
\end{equation}
This relationship is consistent with Eq. 1 of Ref. \cite{SK94}, but
has a more compact and easily accessible form. It is valid for
arbitrary truncation number $N$. By taking $N\rightarrow \infty$, we
have an alternative proof of the isospectrality of these two cavities,
which uses tools more familiar to the physicist. Clearly, the two
cavities represent the same quantum problem, but in different
representations.

\section{Results and discussion}
Since our computations yield exactly the same energies for the two
cavities, only one spectrum is presented. There is a subset of levels,
those of the unit triangle, which are analytically known, and these
have to be added to the zeros of $\det(M_1)$.  Introducing an energy
unit $E_u=(\pi/d)^2$, the eigen-energies of a unit square are
$(n_x^2+n_y^2)E_u$. For a unit triangle, due to the antisymmetrization
requirement, the allowed eigen-energies are the same with the
restriction $ n_x>n_y $, with corresponding wavefunction
\begin{equation}
\Psi_{n_x,n_y}=\sin\frac{n_x\pi x}{d} \sin\frac{n_y\pi y}{d}
 -\sin\frac{n_x\pi y}{d} \sin\frac{n_y\pi x}{d}
\end{equation}
The first two such states are at $5E_u$, and $10E_u$ which correspond
to modes $(n_x, n_y) = (2,1)$ and $(3,1)$. Comparing with the
experimental or calculated spectrum, they are the $9th$ and $21st$
states. In Fig. 2 we plot these two states, revealing their triangular
nature. Because these states are analytically known, and are observed
experimentally, they can be used to calibrate the experiment.

There are degenerate states within the set of triangular states.  The
first two-fold degenerate pair appears at $E=65E_u$ with modes (7,4)
and (8,1). Aside from the triangular states, numerical results show
there are no other degenerate states up to the 600th level.

Table 1 lists 25 levels from the experiment of Ref. \cite{SK94} and
from our calculation.  Column 1 is simply the state sequential
number.  All the energies are in units of $E_u$. The second and the
third columns (Cavity 1 and Cavity 2) are obtained from the measured
frequencies of Table 1 of Ref. \cite{SK94} with the conversion formula
$E/E_u=(2fd/c)^2=0.25842(f/GHz)^2$. If the experiment were done with
air in the cavity as the normal condition, the conversion factor would
be multiplied by 1.0006. Table 1 assumes this situation. These two
sets of data agree with an average percentage error $0.2\%$.

The fourth column (F.D.) is the result from finite differences.
Calculations for grid sizes $ h =d/30$, $\, d/40$, $\, d/50$, were
found to vary quadratically with grid size $h$. Richardson
extrapolation to the limit gives accurate eigenvalues.

The last column (M.M.) is the result from our mode matching
method. Results were found to vary linearly in the variable
$1/N$. Values from $N=50$ and $N=60$ were extrapolated linearly to
$N\rightarrow\infty$.  Comparing F.D. with M.M., the error is
$0.02\%$. This small error probably comes primarily from the finite
difference method.  Comparing the mode matching method with either
cavity's experimental data shows the difference is $0.3\%$ on
average.  Thus the two computational results agree by one order of
magnitude better than the agreement between theory and
experiment. Fig. 3 shows the percentage error relative to the mode
matching method ($E/E_{M.M.}-1$). It reveals both statistical and
systematic error in the data. It should be noticed that due to the
upward trend in the data, an overall scaling to the energy, which may
account for overall dimensional error, will not dramatically reduce
the experimental-theoretical difference. It would be interesting to
understand the source of the difference between theory and the data.

One of the highlighted points of Ref. \cite{SK94} was the ability to
measure the wavefunction. Fig. 4 plots the 1st, 3rd and the 6th
wavefunctions which show qualitative agreement with experiment.  Our
more accurate wavefunctions may be useful in further study on the
problem of classical-quantum correspondence.

We computed the level statistics for the first 600 states as shown in
Fig. 5. The actual spectrum was unfolded using the average density of
states obtained from Weyl's formula as given by Ref. \cite{SK94}. One
sees excellent agreement with the quantum spectrum. Although both
cavities are pseudo-integrable systems\cite{Richens}, the statistics
are closer to GOE than to Poisson. As we have pointed out, there are
degeneracies among the eigenstates of the unit triangle.  This part of
the spectrum is fully integrable. Separating them from the whole
spectrum, the level statistics of the remainder is now very much like
that of GOE, a type of level statistics often associated with
non-integrable systems\cite{Bohigas}.

In summary, we have presented numerical calculations for the
isospectral problem in domains constructed from right angle
$45^{\circ}$ triangles. We found accurate theoretical results which
confirm the experiment and give an absolute reference to the data.  We
point out a subset of eigen-states which are analytically solvable and
can be used the calibrate the experiment.  The mode matching method
also yields a simple analytical proof of isospectrality.

\appendix
\section{Mode matching formula for cavity 2}
The auxiliary boundary values are expanded as:
 \begin{eqnarray}
\Psi_{AB}&=&\sum_n A_{2n} \sin\frac{n\pi x}{d}\quad,\\
\Psi_{BC}&=&\sum_n B_{2n} \sin\frac{n\pi (y-d)}{d}\quad,\\
\Psi_{CD}&=&\sum_n C_{2n} \sin\frac{n\pi (2d-y)}{d}\quad,\\
\Psi_{CE}&=&\sum_n D_{2n} \sin\frac{n\pi (x-d)}{d}\quad.
\end{eqnarray}
Then:
\FL
 \begin{equation}
\Psi_A=\sum_n A_{2n}\sin\frac{n\pi x}{d}\,
\frac{\sin\alpha_n(3d-y)}{\sin\alpha_n d}\quad,
\end{equation}

\FL
 \begin{eqnarray}
\Psi_B&=&\sum A_{2n}\Bigl[\sin\frac{n\pi x}{d}\,
\frac{\sin\alpha_n(y-d)}{\sin\alpha_n d}\nonumber\\
&&\phantom{\sum A_{2n}}
-\sin\frac{n\pi(2d-y)}{d}\,
\frac{\sin\alpha_n(d-x)}{\sin\alpha_n d}\Bigr]\nonumber\\
&+&\sum_n B_{2n}\Bigl[\sin\frac{\sin n\pi(y-d)}{d}\,
\frac{\sin\alpha_n x}{\sin\alpha_n d}\nonumber\\
&&\phantom{\sum B_{2n}}
-\sin\frac{n\pi(d-x)}{d}\,\frac{\sin\alpha_n(2d-y)}{\sin\alpha_n
d}\Bigl]
\quad,
\end{eqnarray}

\FL
 \begin{eqnarray}
\Psi_C&=&\sum_n B_{2n} \sin \frac{n\pi(y-d)}{d}\,
\frac{\sin\alpha_n(2d-x)}{\sin\alpha_n d}\nonumber\\
  &+& \sum_n C_{2n}\sin \frac{n\pi(2d-y)}{d}\,
 \frac{\sin\alpha_n(x-d)}{\sin\alpha_nd}\nonumber\\
  &+& \sum_n D_{2n}\sin\frac{n\pi(x-d)}{d}\,
\frac{\sin\alpha_n(2d-y)}{\sin\alpha_n d}\quad,
\end{eqnarray}

\FL
 \begin{eqnarray}
\Psi_D&=&\sum_n C_{2n}\Bigl[\sin \frac{n\pi(2d-y)}{d}\,
\frac{\sin\alpha_n(3d-x)}{\sin\alpha_nd}\nonumber\\
&&\phantom{\sum C_{2n}}
  -\sin\frac{n\pi(x-2d)}{d}\,
\frac{ \sin\alpha_n(y-d)}{\sin\alpha_nd}\Bigr]\quad,
\end{eqnarray}

\FL
 \begin{eqnarray}
\Psi_E&=&\sum_n D_{2n}\Bigl[\sin \frac{n\pi(x-d)}{d}\,
\frac{\sin\alpha_ny}{\sin\alpha_nd}\nonumber\\
&&\phantom{\sum D_{2n}}
  -\sin\frac{n\pi(d-y)}{d}\,
\frac{ \sin\alpha_n(2d-x)}{\sin\alpha_nd}\Bigr]\quad,
\end{eqnarray}

Matching normal derivatives:
 \begin{eqnarray}
\int_0^d\sin\frac{m\pi x}{d}
\Bigl[\frac{\partial \Psi_B}{\partial y}
     -\frac{\partial \Psi_A}{\partial y}
\Bigr]_{y=2d}&=&0\label{eq:mm1}\\
\int_d^{2d}\sin\frac{m\pi (y-d)}{d}
\Bigl[\frac{\partial \Psi_B}{\partial x}
     -\frac{\partial \Psi_C}{\partial x}
\Bigr]_{x=d}&=&0\\
\int_d^{2d}\sin\frac{m\pi (y-d)}{d}
\Bigl[\frac{\partial \Psi_C}{\partial x}
     -\frac{\partial \Psi_D}{\partial x}
\Bigr]_{x=2d}&=&0\\
\int_d^{2d}\sin\frac{m\pi (x-d)}{d}
\Bigl[\frac{\partial \Psi_C}{\partial y}
     -\frac{\partial \Psi_E}{\partial y}
\Bigr]_{y=d}&=&0\label{eq:mm4}
\end{eqnarray}
Allowing $m$ to run from 1 to $N$, Eqs. (\ref{eq:mm1})--(\ref{eq:mm4})
give rise to Eq. (\ref{eq:M2}).

\begin{table}
\caption{The first 25 eigen-values from the experiment of
Ref. \protect\cite{SK94} and from our numerical calculations.}
\begin{tabular}{rrrrr}
 n& Cavity 1 & Cavity 2 &F. D.\phantom{A}&M. M.\phantom{A}\\
\tableline
  1 & 1.02471 & 1.02481  & 1.02893 58&  1.02853 50\\
  2 & 1.46899 & 1.47194  & 1.48186 54&  1.48146 72\\
  3 & 2.08738 & 2.08831  & 2.09824 94&  2.09746 74\\
  4 & 2.64079 & 2.63985  & 2.64971 54&  2.64954 66\\
  5 & 2.93297 & 2.92949  & 2.93817 62&  2.93743 35\\
  6 & 3.72675 & 3.71892  & 3.73268 94&  3.73233 41\\
  7 & 4.28393 & 4.28388  & 4.29519 27&  4.29472 78\\
  8 & 4.67021 & 4.66917  & 4.67766 52&  4.67753 22\\
  9 & 4.98838 & 4.98531  & 5.00000 19&  5.00000 00\\
 10 & 5.27908 & 5.27278  & 5.29147 53&  5.29027 51\\
 11 & 5.78755 & 5.78371  & 5.80153 08&  5.80113 84\\
 12 & 6.41357 & 6.43781  & 6.43389 42&  6.43215 56\\
 13 & 6.84891 & 6.84718  & 6.86626 01&  6.86622 62\\
 14 & 7.15242 & 7.16045  & 7.15980 24&  7.15934 32\\
 15 & 7.67783 & 7.70604  & 7.69473 74&  7.69241 71\\
 16 & 8.44285 & 8.45947  & 8.46365 45&  8.46325 68\\
 17 & 8.57859 & 8.62220  & 8.61353 59&  8.61116 89\\
 18 & 8.99495 & 8.97209  & 9.01240 54&  9.01034 93\\
 19 & 9.60312 & 9.59562  & 9.60996 82&  9.60979 08\\
 20 & 9.92583 & 9.93689  & 9.92113 11&  9.92103 96\\
 21 &10.00330 &10.03932  &10.00000 76& 10.00000 00\\
 22 &10.55227 &10.55740  &10.57102 01& 10.56973 65\\
 23 &11.09578 &11.10035  &11.06691 65& 11.06572 72\\
 24 &11.41874 &11.40569  &11.41955 09& 11.41884 99\\
 25 &11.99364 &11.98033  &11.98464 97& 11.98408 03
\end{tabular}
\end{table}

\begin{figure}
\caption{Geometric shapes of the two two-dimensional domains with
identical spectra. The dotted lines are boundary lines dividing
the cavities into smaller regions.
}
\end{figure}

\begin{figure}
\caption{Contour plot of wavefunctions for the 9th (the top row) and
the 21st (bottom row) states which are eigen-states of a closed unit
triangle. The 21st state is exactly the doubling of the 9th state.
}
\end{figure}

\begin{figure}
\caption{Percentage error relative to the mode matching method.
Finite difference method (stars); experimental values: Cavity 1
(triangles), cavity 2 (squares).}
\end{figure}

\begin{figure}
\caption{Contour plot of the wavefunctions for states 1, 3, and
6, which correspond to the measured wavefunctions in Fig.
2 of Ref. \protect\cite{SK94}. }
\end{figure}

\begin{figure}
\caption{Nearest level spacing distribution $P(X)$ and $\Delta_3$
statistics of the spectrum. The upper panel is for the full set of the
first 600 states. The bottom panel is after the triangular states
were removed. Also plotted are Wigner and Poisson distribution
curves.
}
\end{figure}


\begin{references}

\bibitem{K66} M. Kac, Am. Math. Monthly, {\bf 73} (1966) 1.

\bibitem{GWW92} C. Gordon, D. Webb and S. Wolpert, Invent. Math.
{\bf 110} (1992) 1-22

\bibitem{SK94} S. Sridhar and A. Kudrolli, Phys. Rev. Lett.
{\bf 72} (1994) 2175-8.

\bibitem{Richens} P. J. Richens and M. V. Berry, Physica {\bf 2D}, 495 (1981).

\bibitem{Bohigas} O. Bohigas, M. J. Giannoni, and C. Schmit, Phys.
Rev. Lett. {\bf 52}, 1 (1984).
\end{references}
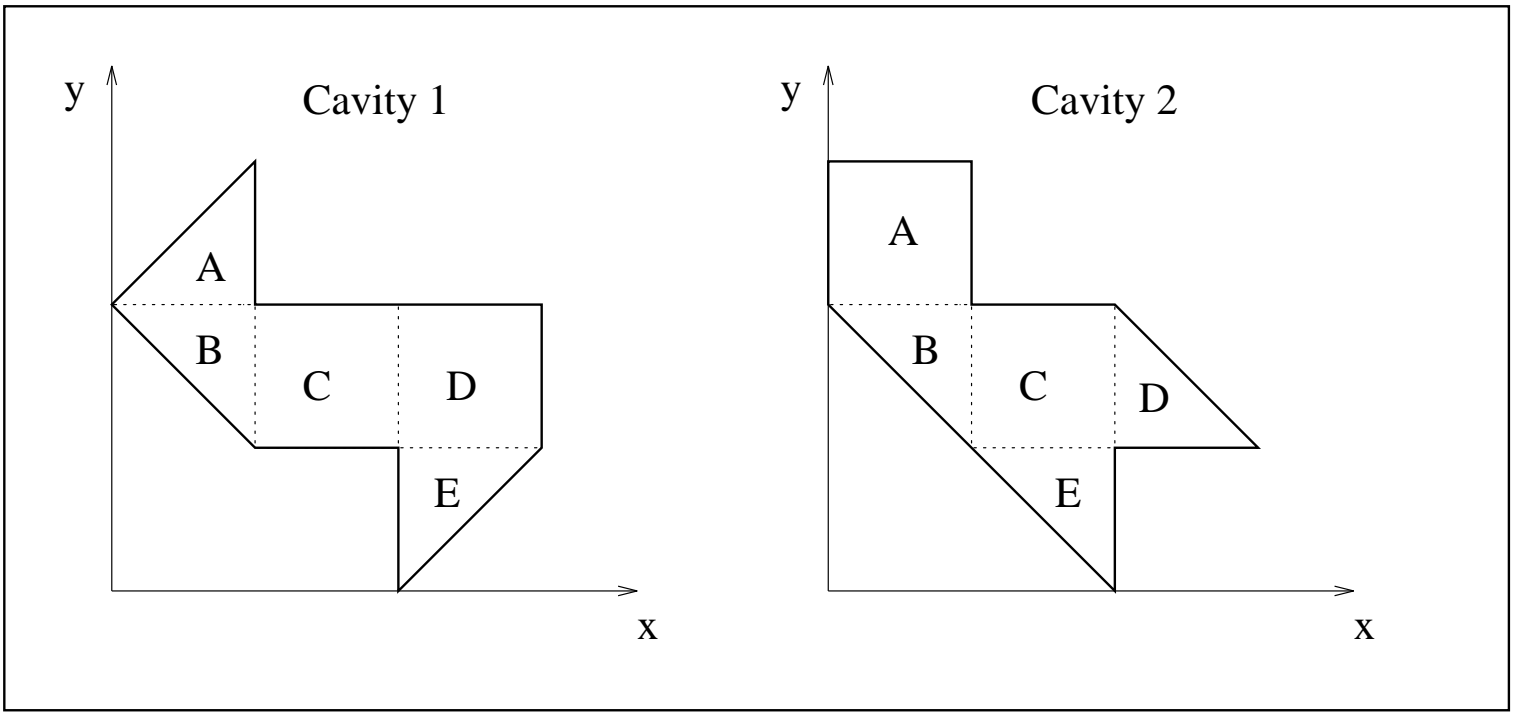
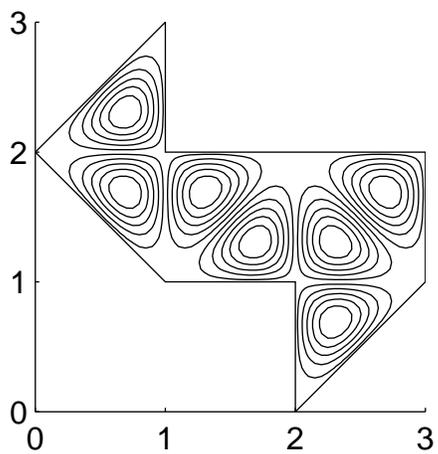
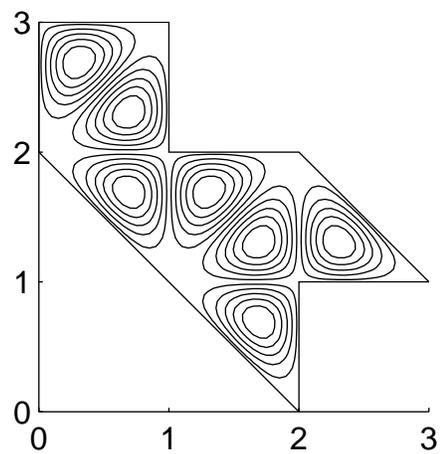
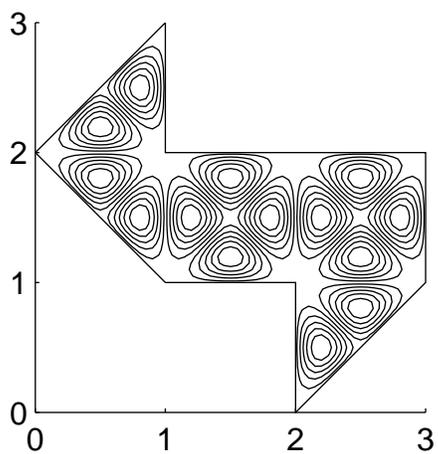
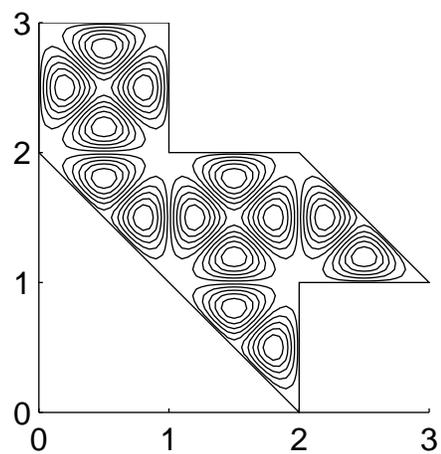
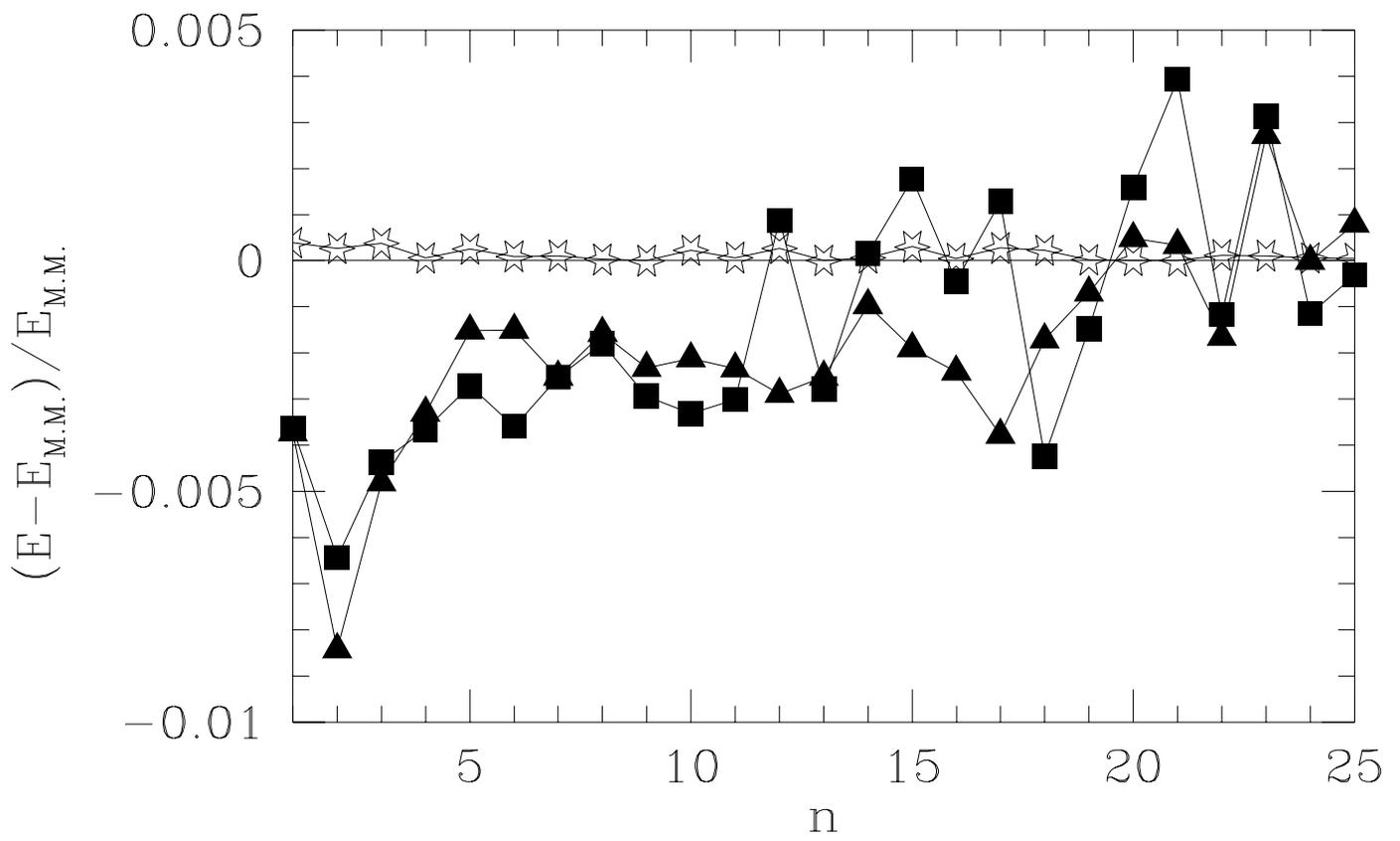
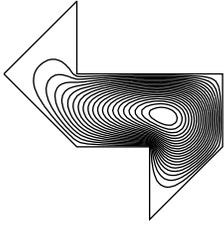
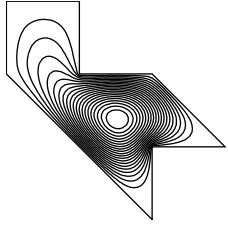
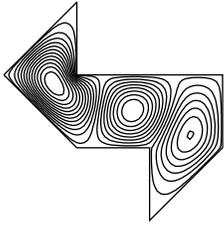
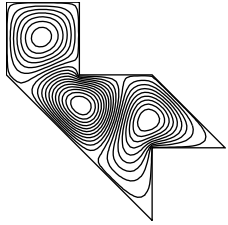
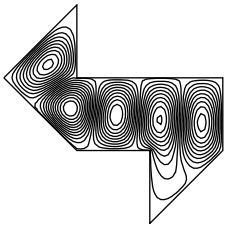
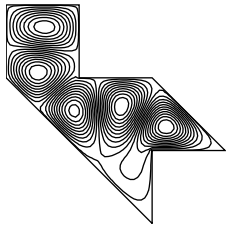
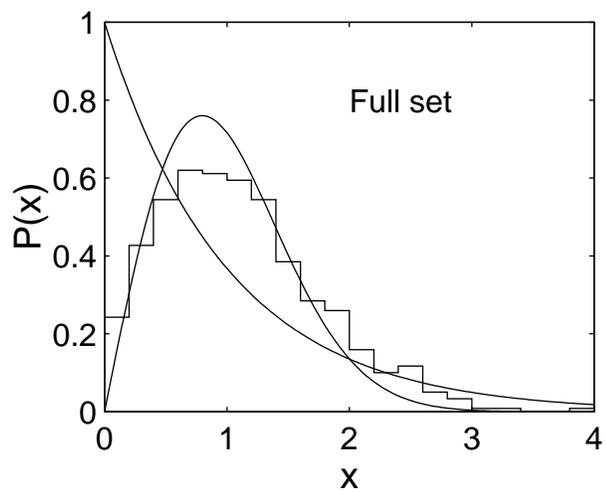
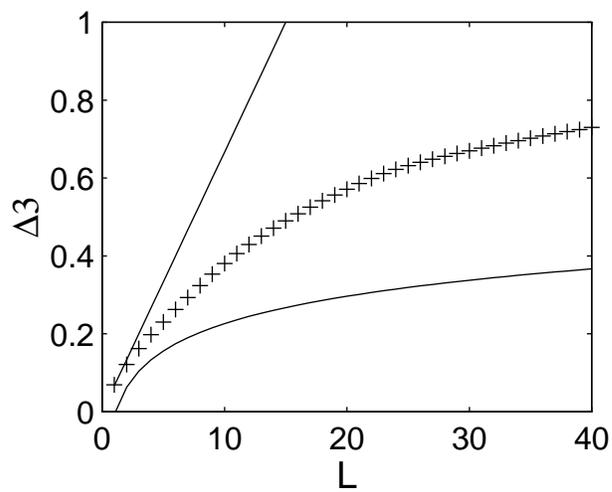
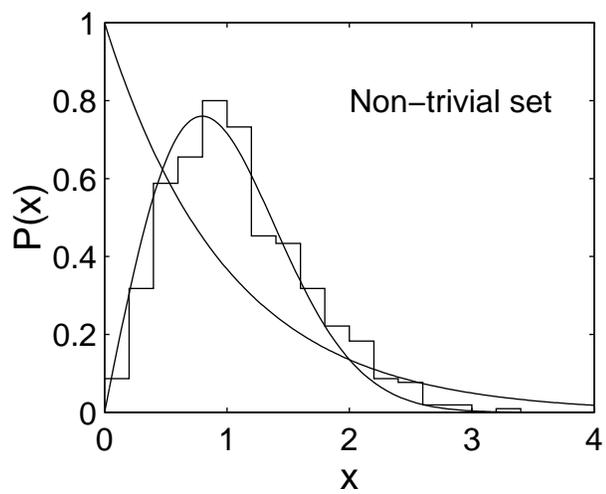
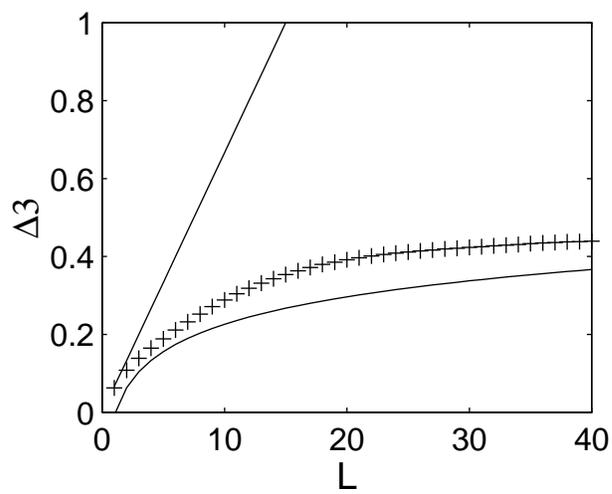
\end{document}